\documentclass[doublecol]{epl2}

\usepackage{amssymb}
\usepackage{amsmath}

\newcommand {\sech} {\textrm{sech}}
\newcommand{\PT}{{$\cal PT$}}

\newcommand{\RE}{\mbox{Re}}

\newcommand{\s}{\sigma}
\newcommand{\lm}{\lambda}
\newcommand{\lmo}{{\lambda_{0}}}
\newcommand{\so}{{\sigma_{0}}}

\newcommand{\tw}{\tilde{w}}
\newcommand{\p}{\partial}

\title{Stability of localized modes in \PT-symmetric nonlinear potentials}
\shorttitle{Localized modes in \PT-symmetric nonlinear potentials} 

\author{D. A.  Zezyulin\inst{1}\thanks{E-mail: \email{zezyulin@cii.fc.ul.pt}} \and Y. V. Kartashov\inst{2} \and V. V. Konotop\inst{1}}
\shortauthor{D. A. Zezyulin  \etal}

\institute{
  \inst{1} Centro de F\'isica Te\'orica e Computacional and Departamento de F\'isica, Faculdade de Ci\^encias, Universidade de Lisboa,   Avenida Professor Gama Pinto 2, Lisboa 1649-003, Portugal\\
  \inst{2} ICFO-Institut de Ciencies Fotoniques, and Universitat Politecnica de Catalunya, Mediterranean Technology Park, 08860 Castelldefels (Barcelona), Spain}

\abstract{We report on detailed investigation  of the stability of localized
modes in the nonlinear Schr\"odinger equations with a nonlinear
parity-time (alias \PT) symmetric potential. We are particularly
focusing on the case where the spatially-dependent  nonlinearity
is purely imaginary. We compute the Evans function of the linear
operator determining the linear stability of localized  modes. Results of the Evans function analysis predict that for sufficiently small dissipation localized modes become stable  when
the propagation constant exceeds certain threshold value. This is
the case for periodic and $\tanh$-shaped complex potentials where
the modes having widths comparable with or smaller than the
characteristic width of the complex potential are stable, while
broad modes are unstable. In contrast, in complex potentials that
change linearly with transverse coordinate all modes are stable,
what suggests that the relation between width of the modes and
spatial size of the complex potential define the stability in the
general case. These results were  confirmed using the direct
propagation of the solutions for the mentioned examples.}

\begin{document}

\maketitle

\section{Introduction}

Since the introduction of the concept of parity-time-
($\mathcal{PT}$-) symmetric potentials~\cite{Bender_first},
non-Hermitian Hamiltonians  possessing purely real spectrum have
received considerable  attention~\cite{Special_issues} due to
their relevance to the quantum mechanics and optics. In the
context of optical applications, it was natural that the concept
was generalized to the  nonlinear systems~\cite{Christodoul},
where the existence of localized modes was shown to be possible
{(we notice that experimental observation of the \PT symmetry in
linear optics was recently reported~\cite{Kipp}).} As a natural
extension of this activity, the existence of stable localized
nonlinear modes in nonlinear \PT-symmetric lattices has been
recently demonstrated in~\cite{AbdKartKonZez}. {Nonlinear gain and
loss compensating each other were also addressed recently within
the framework of the nonlinear dimer model \cite{MirMalKiv}.} A
general important property of nonlinear \PT-symmetric systems is
that they admit continuous families of localized modes
parameterized by the propagation constant, similarly to what
happens for the nonlinear Hamiltonian systems. This is in spite of
the need to  satisfy the balance between dissipation and gain, as
this happens in  dissipative systems of general type  where
localized modes appear as attractors, rather than elements of a
family of solutions. In this sense, the models with \PT-symmetric
potentials occupy  a special ``place'' between the Hamiltonian and
dissipative systems.

A striking effect related to the existence and stability of the
localized modes in nonlinear \PT-lattices is that they can be
stable even in the absence of  modulation of conservative part of
nonlinearity~\cite{AbdKartKonZez}. The  stability, however depends
on the relation between the mode width and the period of the
potential. More specifically only sufficiently narrow modes were
found to be stable. Now we observe, that in the limit when the
width of a mode goes to zero and the potential is a smooth
function, the behavior of the imaginary part of the potential can
be approximated by the respective linear function of the
coordinate (recall, that the imaginary part is an odd function).
For instance, the purely imaginary 
nonlinear \PT-potential $i\sin(2\eta)$ considered
in~\cite{AbdKartKonZez} can be approximated as $i\sin(2\eta)\sim
2i\eta$ for sufficiently narrow modes strongly localized about
$\eta=0$. This leads to the natural question: is it possible to
obtain stable localized modes in a nonlinear \PT-symmetric
potential of a general type, provided the widths of the modes are
small enough? {Here we give an affirmative answer to this question
and illustrate stable modes for three qualitatively different case
examples: of a periodically varying dissipation and gain, i.e. the
lattice case, of dissipation and gain tending to constants at the
infinity (see (\ref{tanh}) below), and to the linearly increasing
gain and losses (the model (\ref{2eta})).}

The aim of the present paper is the analytic study of the linear
stability and dynamics of solitons in nonlinear \PT-symmetric
potentials, giving positive answer to the above question. The main
model we are interested in is the  complex  nonlinear
Schr\"odinger  equation ~\cite{AbdKartKonZez}:
\begin{eqnarray}
\label{CNLS}
i \p_\xi q =-\frac 12 \p^2_\eta q -\left[1 +V(\eta)+ i W(\eta)\right]|q|^2q
\end{eqnarray}
In the optical applications $q$ is the
dimensionless electric field propagating along the $\xi$-direction
($\xi>0$)  with $\eta$ ($\eta\in \mathbb{R}$) being the
transverse coordinate. {We  notice
that the physical mechanism for both  gain and loss  are well
known. The former one refers to the standard two-photon
absorption, which becomes dominating mechanism of losses in semiconductors at sufficiently high intensities, while nonlinear amplification  can be realized in
electrically-pumped semiconductor optical amplifiers (see
e.g.~\cite{Lederer}).} We are interested in
the special case of a \PT-symmetric nonlinear potential where the
  $V(\eta)$ and $W(\eta)$ are both real and obeying the
relations:
\begin{eqnarray}
\label{eq:NPT} V(\eta)=V(-\eta) \quad \mbox{and}\quad
W(\eta)=-W(-\eta).
\end{eqnarray}

\section{Localized modes}
\label{sec:stationary} We look for  
stationary localized solutions of
eqs.~(\ref{CNLS})--(\ref{eq:NPT}), which can be searched in the
form $q(\xi, \eta)=w(\eta)e^{ib\xi}$ subject to the boundary conditions $\lim_{\eta\to\pm\infty}|q(\xi, \eta)|=0$. Bearing in mind optical applications, we refer to $b$ as to the propagation constant. 
The stationary wave function $w(\eta)$ obeys the  equation
\begin{eqnarray}
\label{eq:stat} \frac12\p^2_\eta w-bw+ [1+ V(\eta) +
iW(\eta)]|w|^2w=0.
\end{eqnarray}

Let us calculate how many parameters   one has to introduce  in
order to unambiguously define a localized mode $w(\eta)$.  To this
end, let us first agree that we identify  the modes $w(\eta)$ and
$w(\eta)e^{i\varphi}$, $\varphi\in \mathbb{R}$, which are not
distinguishable from physical point of view. Then a specific
symmetry of eq.~(\ref{eq:stat}) induced by
relations~(\ref{eq:NPT}) suggests that without loss of generality
the localized mode $w(\eta)= w_r(\eta) + iw_i(\eta)$ can be
chosen to be \PT-symmetric, i.e. having  even real part and odd
imaginary one:
\begin{eqnarray}
\label{eq:wrwi} w_r(\eta) = w_r(-\eta), \qquad w_i(\eta)
 =-w_i(-\eta).
\end{eqnarray}

Let  $\tilde{w}(\eta)$ be  some solution of eq.~(\ref{eq:stat})
vanishing as $\eta\to+\infty$, but not
necessarily vanishing as $\eta\to-\infty$. 
Then for $\eta\to+\infty$  the nonlinear term in
eq.~(\ref{eq:stat}) is negligible and  in the corresponding limit
$\tw(\eta)$ is described by the linear equation $\frac12 \p^2_\eta
\tw - b\tw = 0$. Thus   for  $\eta \to+\infty$ the solution
$\tw(\eta)$  behaves as $ \tw(\eta) = Ce^{i\varphi}
[e^{-\sqrt{2b}\eta} + o(1)]$, where $C$ and $\varphi$ are real
constants. For a generic solution $\tw(\eta)$ the constant
$\varphi$ is not   zero, but let us  temporarily restrict
ourselves to the case  $\varphi=0$. Eqs.~(\ref{eq:wrwi}) dictate
that   if  the solution $\tw(\eta)$
 represents a localized mode,  then $\tw(\eta)$  must obey
$\p_\eta\tw_r(0) = 0$ and $\tw_i(0)= 0$. Now, let us  admit
nonzero $\varphi$  in the asymptotics for $\tw(\eta)$. Obviously,
this leads just to multiplication of $\tw(\eta)$ by the factor
$e^{i\varphi}$. Therefore,  we can formulate  a weaker condition
for $\tw(\eta)$ to represent a localized mode: there must exist
such $\varphi$ that $\tw_{i}(0)\cos\varphi + \tw_{r}(0)\sin\varphi
= 0$ and   $\p_\eta \tw_{r}(0)\cos\varphi - \p_\eta
\tw_{i}(0)\sin\varphi = 0$. These equations are compatible if and
only if
\begin{equation}
\label{eq:PTsh}%
 \tw_{r}(0)\p_\eta \tw_{r}(0) + \tw_{i}(0) \p_\eta \tw_{i}(0) =
0.
\end{equation}
Thus any localized mode $w(\eta)$ can be identified with a
solution of the eq.~(\ref{eq:PTsh})  which contains two real
unknowns: $C$ and $b$. If we fix one of them, typically this is
the propagation constant $b$, then eq.~(\ref{eq:PTsh}) results in
one or several solutions for the parameter $C$ what indicates that
eq.~(\ref{eq:stat}) admits \textit{continuous families of
localized modes for fixed $V(\eta)$ and $W(\eta)$}. This feature
is typical for \PT-symmetric (linear or nonlinear) potentials and
constitutes significant difference compared to the conventional
dissipative systems.

\begin{figure}
\begin{center}%
\includegraphics[width=0.8\columnwidth]{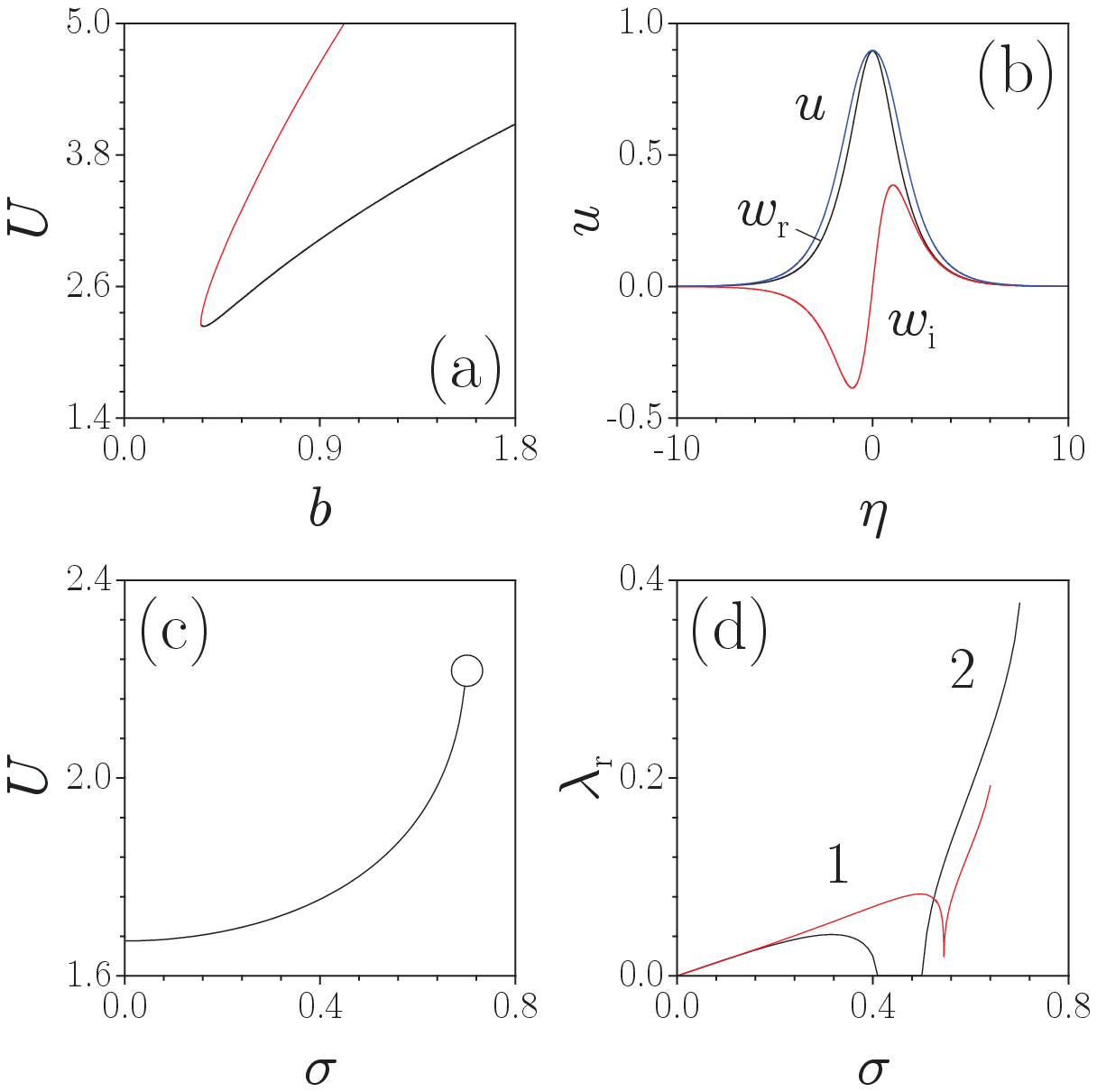}
\end{center}
\vspace{-0.9cm} \caption{Properties of the modes for the potential~(\ref{tanh}).
(a) \textit{Families} of the  \textit{fundamental} and
\textit{higher}  localized modes for $\s= 0.7$. (b) Real and
imaginary parts $w_{r,i}$ as well as the modulus $u=|w|$ of the
fundamental mode at $b=0.35$, $\s= 0.7$.
 (c) The \textit{branch} of the fundamental modes  found for $b = 0.35$. The circle corresponds to
the mode shown in panel (b). (d) Real part $\lambda_r=\RE\lambda$
of perturbation growth rate {\it vs} $\s$ for $b = 0.2$ (curve 1)
and $b = 0.35$ (curve 2).} \label{fig-1}
\end{figure}

In fig.~\ref{fig-1}~(a) we show two families of localized modes
plane $(b, U)$, where  $U=\int|q|^2d\eta$ is the energy flow (the
integration limits are omitted wherever the integration is over
whole real axis), obtained  for the nonlinear potential
\begin{eqnarray}
\label{tanh} V(\eta) =  0, \quad\mbox{and}\quad W(\eta)=  \sigma\tanh(2\eta)
\end{eqnarray}
with $\s=0.7$.  The modes exist only if $b$ exceeds certain
threshold value. The modes of the lower (upper) curve in
fig.~\ref{fig-1}~(a) can be referred  to    as
\textit{fundamental} (\textit{higher}) modes.  A typical profile
of a fundamental mode is shown in fig.~\ref{fig-1}~(b). Below we
focus on the fundamental modes.

On the other hand, the localized modes can be considered
as bifurcating from the limit $\s=0$, where eq.~(\ref{CNLS})
reduces to the conventional  nonlinear Schr\"odinger equation. In
fig.~1~(c) we show a \textit{branch} of   fundamental modes     on the plane $(\s,
U)$  found for a fixed value of $b$. %
The continuation
 from the limit $\s=0$ will be used below  as an approach for analytical
investigation of stability of the fundamental modes.

\section{Linear stability analysis}
\label{sec:stabil} Substituting the perturbed solution $q(\xi,
\eta) = e^{ib\xi}\left[w(\eta) + {e^{\lm \xi}p_+ + e^{\bar{\lm}
\xi}\bar{p}_-}\right]$, where $|p_\pm|\ll |w|$, and  the overline
stands for the complex conjugation, in eq.~(\ref{CNLS}) and
linearizing it around $w(\eta)$ one arrives at  the eigenvalue
problem  $L \vect{p} =\lambda \vect{p}$, where $\vect{p} =
{(p_++p_-, i(p_-  - p_+))}^T$ (hereafter the superscript $T$
stands for  matrix transposition) and  $L$ is given by
\begin{eqnarray}
\label{L}
L = \left( \begin{array}{cc}  %
N_{11}  &  - \frac12 \p^2_\eta  + N_{12}\\
\frac12  \p^2_\eta + N_{21}  &  N_{22}%
\end{array}
\right),
\end{eqnarray}
where
\begin{subequations}
\label{eq:Ljk1}
\begin{eqnarray}
N_{11} = -2 [1 + V(\eta)]w_rw_i  - W(\eta)[3w_r^2 + w_i^2],
\\
N_{22} = 2 [1 + V(\eta)]w_rw_i  -  W(\eta)[w_r^2  +  3w_i^2],
\\%
N_{12} =    b - [1 + V(\eta)][w_r^2 + 3w_i^2] - 2W(\eta) w_rw_i,
\\
N_{21} =   - b + [1 + V(\eta)][3w_r^2 + w_i^2]  - 2W(\eta) w_rw_i.
\label{eq:Ljkend}
\end{eqnarray}
\end{subequations}
The    mode $w(\eta)$ is unstable if and only if there exists an
eigenvalue $\lambda$   with positive real part.

For  $V(\eta) = W(\eta)\equiv 0$    the localized mode  $w(\eta)$
is a standard NLS soliton: $w_r(\eta) = w_r^{(0)} =\sqrt{2b}\,
\sech(\sqrt{2b}\eta)$ and $w_i(\eta) = w_i^{(0)} \equiv 0$.
Designating the operator $L$ in this case by $L^{(0)}$, we recall
that  the spectrum of  $L^{(0)}$ is well
known~{\cite{KuzRubZakh}}. In particular, the point
spectrum of $L^{(0)}$ consists of the only eigenvalue
$\lambda_0=0$  which is isolated and  has algebraic multiplicity
(a.m.) equal to 4 and geometric multiplicity (g.m.) equal to 2.
The eigenfunctions corresponding to $\lm_0$ read
\begin{equation}
\label{eq:psi11psi12} \vect{\psi}^{(0)}_{11} =  \left( \p_\eta
w_{r}^{(0)} , 0\right)^T, \quad   \vect{\psi}^{(0)}_{12} =
\left(0, w_r^{(0)}\right)^T.
\end{equation}
There also exist  two generalized eigenfunctions, namely 
\begin{equation}
\label{eq:psi21psi22} \vect{\psi}_{21}^{(0)} =  \left(0, -\eta
w_r^{(0)}\right)^T  \mbox{and }   \vect{\psi}_{22}^{(0)} =
\left(\p_b w_r^{(0)}, 0\right)^T,
\end{equation}
 such that $L\vect{\psi}^{(0)}_{2j} = \vect{\psi}^{(0)}_{1 j}$, $j=1, 2$.  Here $\p_b w_r^{(0)}$ is obtained by means of  differentiation of $w_r^{(0)}$ with respect to $b$.

Let us now assume that $V(\eta)$  and $W(\eta)$ are not equal to
zero but    satisfy \PT-symmetry relations~(\ref{eq:NPT}).  If at
the same time $V(\eta)$ and $W(\eta)$ are small enough, then  they
can be considered as   a perturbation to the operator $L^{(0)}$.
Behavior of the spectrum of $L^{(0)}$ subject to the perturbation
determines linear stability of the localized mode.  In particular,
it is crucial to understand what happens to the multiple
eigenvalue {$\lambda=0$} since a generic  perturbation of
the operator $L^{(0)}$ leads to splitting of {the
 eigenvalue $\lambda=0$} into several simple eigenvalues.
As a result, unstable eigenvalues can arise in the vicinity of
{$\lambda=0$}.

\begin{figure}
\includegraphics[width=\columnwidth]{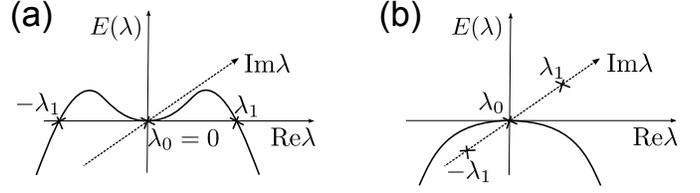} \caption{Possible
behavior of the Evans function $E(\lambda)$ for real $\lambda$ is
schematically illustrated.   (a) $\p^2_{\lmo}E
>0$ and apart from the zero $\lmo=0$ the Evans function has two
real zeros $\pm\lm_1$.   (b) $\p^2_{\lmo}E < 0$ and no real zeros
of the Evans function exists besides of {$\lambda_0=0$}. Instead,
there are two purely imaginary zeros $\pm\lm_1$.} \label{fig-2}
\end{figure}

The multiplicity of the eigenvalue {$\lambda=0$}  is related to
rotational (i.e. phase) and translational symmetries of the model.
The dissipative perturbation introduced by $V(\eta)$ and $W(\eta)$
breaks the translational symmetry and preserves the rotational
one. Due to the last fact {$\lambda=0$} remains an eigenvalue for
$L$. However $\vect{\psi}^{(0)}_{11}$ ceases to be the
eigenfunction corresponding to {$\lm=0$} and the only
eigenfunction for {$\lambda=0$} is given by $ \vect{\psi}_{12} =
(- w_i, w_r)^T$. Disregarding parity of the functions $w_{r,
i}(\eta)$  we  observe that if $\lambda$ is an eigenvalue of the
operator $L$ then $\bar{\lambda}$  is also an eigenvalue. Then,
recalling eqs.~(\ref{eq:wrwi}) we find that  $-\lambda$ is an
eigenvalue, as well. Therefore  upon the perturbation  exactly two
simple eigenvalues arise in the vicinity of {$\lm=0$}  and those
eigenvalues have opposite signs and  either purely real or purely
imaginary (see  fig.~\ref{fig-2}).

\section{Evans function for \PT-symmetric potentials}
The operator $L$ can be associated with the Evans function $E(\lambda)$~\cite{GardnerJones, KapEvans}, which is
an analytic function defined on the whole complex plane except
for the points of the continuous spectrum of $L$. An important property of the  Evans
function  is that it has a zero at some $\lambda$ if and only if $\lambda$ is an
eigenvalue of $L$. In addition, the order of that zero is
equal to the a.m. of the eigenvalue $\lambda$.

Before proceeding with   explicit definition of the Evans function
for
 the case at hand, we recall  that perturbation of the operator $L$  leads to that the eigenvalue $\lambda_0$ has  a.m.~$=2$. Hence the Evans function corresponding to the perturbed operator $L$
 has a zero  of the second order at  {$\lambda=0$}: $E({0})
= \p_\lmo E = 0$, $\p^2_\lmo E \ne 0$ (hereafter $\p^j_\lmo$ stays
for $j$th partial derivative with respect to $\lm$ evaluated at
$\lm={0}$).  Without loss of generality one can assume
that $E(\lambda) < 0$ for all $\lambda \gg 1$. Then the stability
of the stationary mode is determined by the  sign of $\p^2_\lmo
E$: if $\p^2_\lmo E > 0$ then $E(\lambda)$ necessarily has exactly
one positive and   one negative zero what corresponds to
instability; {\it vice versa}, if $\p^2_\lmo E < 0$ then  both
roots of  the Evans function lie on the imaginary axis, and  the
solution is stable. These considerations are illustrated in
fig.~\ref{fig-2}

In order to define the Evans function for the operator $L$ given by  eq.~(\ref{L})
 we rewrite the eigenvalue problem  $(L-\lm)\vect{p} = 0$
 in the form of the system of four first-order ODEs $\p_\eta\vect{Y} = M\vect{Y}$, where  $\vect{Y} = (p_r, p_i, \p_\eta p_r, \p_\eta p_i)^T$ and
\begin{equation}
    M = \left(%
    \begin{array}{cccc}%
    0&0&1&0\\
    0&0&0&1\\
    -2N_{21}&2(\lm - N_{22})&0&0
    \\
    2(N_{11}-\lm)&2N_{12}&0&0%
    \end{array}\right)
\end{equation}
The coefficients $N_{jk}$  are given by  eqs.~(\ref{eq:Ljk1}).
Parity of the functions $w_r(\eta)$ and $w_i(\eta)$  imposed by
eqs.~(\ref{eq:wrwi})  implies that $N_{12}$ and $N_{21}$ are even
while  $N_{11}$ and $N_{22}$ are odd as functions of $\eta$.

{For the sake of simplicity  we
 suppose that
 $W(\eta) = \s
W_0(\eta)$, where
$\s\ll 1 $ is the small parameter
 while $V(\eta)=\mathcal{O}(\s^3)$.}
Then the coefficients $N_{jk}$ become also dependent on $\s$.
Obviously, now substitution $\s\to -\s$ is equivalent to
substitution $\eta\to-\eta$ what in turn is equivalent to
$w_r(\eta)\to w_r(\eta)$ and $w_i(\eta)\to -w_i(\eta)$.  As a
result, $N_{12}$ and $N_{21}$ are even  while $N_{11}$ and
$N_{22}$ are odd in $\s$.

Now we choose four solutions $\vect{Y}_j(\eta; \sigma; \lambda)$,
$j=1,\ldots4$,  such that $\vect{Y}_1$ and $\vect{Y}_2$  are
linearly independent in $\eta$ and vanish as $\eta\to-\infty$
while $\vect{Y}_3$ and $\vect{Y}_4$ are linearly independent and
vanish as $\eta\to +\infty$. Then the Evans function is given as
$4\times4$ $\eta$-independent determinant~\cite{GardnerJones}:
$E(\sigma; \lambda) = \det[\vect{Y}_1,  \vect{Y}_3,  \vect{Y}_2,
\vect{Y}_4]$. In order for $E(\s; \lambda)$ to be unambiguously
defined and to depend analytically on $\s$ and $\lambda$, we
explicitly fix the choice of the solutions $\vect{Y_j}$  setting
\begin{eqnarray*}
\vect{Y}_{1,3}(\eta; 0; {0})%
& =& \left(%
\begin{array}{c}
\vect{\psi}^{(0)}_{11}\\%
 {\p_\eta\vect{\psi}^{(0)}_{11}}
\end{array}
\right)=%
\left(\!%
 \p_\eta w_{r}^{(0)}, 0, \p^2_\eta w_{r}^{(0)}, 0\!\right)^T,\\%
\vect{Y}_{2,4}(\eta; 0; {0})%
&= &\left(%
\begin{array}{c}
\vect{\psi}^{(0)}_{12}\\%
{\p_\eta\vect{\psi}^{(0)}_{12}}%
\end{array}
\right)%
=\left(%
0 ,w_r^{(0)},  0,  \p_\eta w_r^{(0)}\right)^T%
\end{eqnarray*}
(recall that $\vect{\psi}^{(0)}_{11}$ and $\vect{\psi}^{(0)}_{12}$
are given by eqs.~(\ref{eq:psi11psi12})). Requiring this choice to
be consistent with parity of the coefficients $N_{jk}$ we find
that
\begin{subequations}
\label{eq:Ysymms1}
\begin{eqnarray}
    \vect{Y}_{3}(\eta; \pm\s; \mp\lm) = \phantom{-}\vect{Y}_1(-\eta; \s;
    \lm)\ast
     \vect{J}_\pm%
    \\
    \vect{Y}_{4}(\eta; \pm\s, \mp\lm) = \pm \vect{Y}_2(-\eta; \s; \lm)\ast
     \vect{J}_\pm
\end{eqnarray}
\end{subequations}
where $\vect{J}_+
    =(-1, \phantom{-}1, \phantom{-}1, -1)^T$, $\vect{J}_-=(-1, {-}1, \phantom{-}1, \phantom{-}1)^T$, and $\ast$ stays for the element-wise multiplication of the matrices.
  Eqs.~(\ref{eq:Ysymms1})   imply  that  the Evans function is even with respect to both its arguments: $E(\s; \lm) = E(\s; -\lm) =  E(-\s; \lm)$. Respectively,  the second derivative  of the Evans function at $\lm = \lm_0$ can be searched in the form of the following expansion:
  \begin{equation}
  \label{eq:Ell}
    \p^2_\lm E(\s; \lm_0) = \sigma^{2k} E_{2k} + o(\s^{2k}), \quad   E_{2k}\ne 0,
   \end{equation}
  where    $k=1, 2,\ldots$ is to be determined.

Subject to the perturbation,  i.e. for $\s\neq 0$,  the
eigenfunction $\vect{\psi}_{11}$ disappears; the solutions
$\vect{Y}_{1, 3}(\eta; \s; {0})$  become unbounded and no longer
correspond to any eigenfunction of the operator $L$. However, the
eigenfunction $\vect{\psi}_{12}$ defined above persists and
depends smoothly on $\s$.  As a result,   for  all  $\s$ the
following equalities hold:
\begin{equation}
\nonumber
 \vect{Y}_{2,4}(\eta; \s; {0})
 = \left(\!\!
\begin{array}{c}
\vect{\psi}_{12}
\\
\p_\eta\vect{\psi}_{12}
\end{array}
\!\!\right)= \left( \! -w_i , w_r
,
-\p_\eta w_i
,
\p_\eta w_r
 \!\right)^T.
\end{equation}
Since $\lm_0$  has a.m.=2 in the spectrum of the perturbed
operator,  Lemma~3.3 from~\cite{KapEvans} can be applied.  It
states that $\p_\lm \vect{Y}_2(\eta; \s; {0}) = \p_\lm
\vect{Y}_4(\eta; \s; {0})$ for all $\eta$ and $\s$.
Bearing in mind  these facts  and  differentiating
straightforwardly the Evans function  one arrives at the
following expression for the second derivative of the Evans
function at $\lambda={0}$:
  \begin{eqnarray}
  \label{eq:Ellwedge}
    \p^2_\lm E(\sigma; {0})   =   \det[\vect{Y}_1,  \vect{Y}_3, \p^2_\lmo (\vect{Y}_2 - \vect{Y}_4),  \vect{Y}_2].
  \end{eqnarray}

Let us now recall that $E(\s; \lm)$ is an even  function of $\s$.
It means that $\p_\so\p^2_\lmo E = 0$ where $\p^j_{\so}$ stands
for the  $j$th partial derivative with respect to $\s$ evaluated
at $\s_0=0$. Calculating the second derivative with respect to
$\s$ and evaluating it at $\s_0$ one arrives at
\begin{eqnarray}
    \label{eq:Essll}
  &&\p^2_\so\p^2_\lmo E  =  \det[\p^2_\so(\vect{Y}_1 - \vect{Y}_3),  \vect{Y}_1, \p_\lmo^2(\vect{Y}_2-\vect{Y}_4),   \vect{Y}_2] +  \nonumber\\%
  &&\det[\p_\so(\vect{Y}_1 - \vect{Y}_3), \vect{Y}_1, \p_\so\p_\lmo^2(\vect{Y}_2-\vect{Y}_4), \vect{Y}_2].
\end{eqnarray}
Using   parity of the coefficients $N_{ij}$ and the symmetries of
the solutions $\vect{Y}_j$ given by eqs.~(\ref{eq:Ysymms1}) one
can  {recognize} that  r.h.s. of eq.~(\ref{eq:Essll}) generically
is not equal to zero, i.e. we can set $k=1$ in eqs.~(\ref{eq:Ell})
and obtain:
\begin{equation}
    \label{eq:EvFunDer}
    \p_\s \p^2_\lm E(0; {0}) = 0, \quad  \p^2_\s \p^2_\lm E(0; {0}) \ne 0.
\end{equation}
{We failed to obtain expressions for the mixed derivatives
$\p^2_\so(\vect{Y}_1 - \vect{Y}_3)$ and
$\p_\so\p_\lmo^2(\vect{Y}_2-\vect{Y}_4)$ which would  allow for
efficient analytical or numerical computation of $\p_\so^2
\p^2_\lmo E(0; {0})$. However, using the information given by
(\ref{eq:EvFunDer}) one can use another representation for
derivatives of the Evans function~\cite{KapEvans} which fits
better for analytical and numerical investigation.}
To this end, we recall that apart from the eigenfunction
$\vect{\psi}_{12}$  there exists a generalized eigenfunction
$\vect{\psi}_{22}$ such that $L\vect{\psi}_{22}=\vect{\psi}_{12}$.
The adjoint operator  for $L$ reads  $L^\dag=L^T$. There also
exists the  adjoint eigenfunction $\vect{v}$ such that $L^\dag
\vect{v} =0$. It is found in~\cite{KapEvans} that for any $\s\ne
0$ the second derivative of the Evans function at {$\lm=0$} can be
found as $\p^2_\lm E(\sigma; {0})=2\langle \vect{\psi}_{21},
\vect{v}\rangle$,
 where $\langle \vect{\cdot}, \vect{\cdot}\rangle$ represents standard $L^2$ inner product of vector-valued functions.
Taking into account  eqs.~(\ref{eq:EvFunDer})   one can  construct
an expansion
  for $\langle \vect{\psi}_{21}, \vect{v}\rangle$  with respect to the small parameter $\s$,   $0<|\s|\ll 1$. To this end, we firstly write down an expansion  for the stationary mode itself.  {Eq.~(\ref{eq:stat}) dictates that  this expansion acquires the following specific
  form:}
$w_r=w_r^{(0)}+\s^2 w_r^{(2)}+o(\s^2)$, $w_i=\s
w_i^{(1)}+o(\s^2)$, where the coefficients $w_r^{(2)}$  and
$w_i^{(1)}$ solve the equations
\begin{subequations}
\begin{eqnarray}
L^{+}w_{r}^{(2)} = - w_r^{(0)}w_i^{(1)}[w_i^{(1)} - W_0(\eta)w_r^{(0)}],\\%
L^- w_{i}^{(1)}  = -W_0(\eta)(w_r^{(0)})^3, \label{eq:Lwi}
\end{eqnarray}
\end{subequations}
and $L^{\pm} = \frac12 \p^2_\eta  - b + (2\pm
1)\left(w_r^{(0)}\right)^2$. Notice that $w_r^{(2)}$ is an even
function of $\eta$  while $w_i^{(1)}$ is odd. Next, {using
eqs.~(\ref{eq:psi21psi22}) and the definition of
$\vect{\psi}_{22}$ (i.e. the equation $L\vect{\psi}_{22} =
\vect{\psi}_{12} $)} we obtain an expansion for the generalized
eigenfunction:
\begin{equation}
\label{eq:psi22} \vect{\psi}_{22}= \left(\begin{array}{c}
\p_b w_{r}^{(0)}\\
0
\end{array}\right) + \s
\left(\begin{array}{c}
0\\
\chi^{(1)}
\end{array}\right) + o(\s),
\end{equation}
where the coefficient $\chi^{(1)}$ is an  odd function of $\eta$
solving the equation
\begin{equation}
\label{L-}
L^-\chi^{(1)} =
  -w_r^{(0)}\p_b w_{r}^{(0)}[3W_0(\eta) w_r^{(0)} + 2 w_i^{(1)}]
 + w_{i}^{(1)}.
\end{equation}
{Using the definition the adjoint eigenfunction $\vect{v}$ and requiring the derivative $\p^2_\lm E(\sigma;
{0})$   to satisfy constrains~(\ref{eq:EvFunDer}) we observe that expansion for  $\vect{v}$ must have the  form}:
\begin{equation}
\label{rhophi}
\vect{v} = \s\left(\!\begin{array}{c}
0\\
\p_\eta w_{r}^{(0)}
\end{array}\!\right) + \s^2
\left(\!\begin{array}{c}
\rho^{(1)}\\
0
\end{array}\!\right) +
\s^3
\left(\!\begin{array}{c}
0\\
\phi^{(2)}
\end{array}\!\right) +
o(\s^3).
\end{equation}
Then   $\rho^{(1)}$ is given as $\rho^{(1)} = \beta w_r^{(0)} +
f$, where $f$  is a particular solution of the equation
\begin{equation}
\label{eq:Lf} L^-f = w_r^{(0)}\p_\eta w_{r}^{(0)} [2w_i^{(1)} -
W_0(\eta)w_r^{(0)}].
\end{equation}
The solvability condition for eq.~(\ref{eq:Lf}) (i.e.
orthogonality of its  r.h.s. to $w_r^{(0)}$) is automatically
provided as long as eq.~(\ref{eq:Lwi}) holds.  At the same time,
the coefficient $\beta$ should be chosen to satisfy the
solvability condition of the equation with respect to
$\phi^{(2)}$:
\begin{eqnarray}
\label{eq:phi2}
L^+\phi^{(2)}  =  [2w_i^{(1)} + 3W_0(\eta)w_r^{(0)}]w_r^{(0)}\rho^{(1)} -
\nonumber \\
  \p_\eta w_{r}^{(0)} [6w_r^{(0)}w_r^{(2)}  + (w_i^{(1)})^2 - 2W_0(\eta) w_r^{(0)}w_i^{(1)}].
\end{eqnarray}
Respectively, we require   r.h.s. of  eq.~(\ref{eq:phi2}) to be
orthogonal to $\p_\eta w^{(0)}_{r}$ what yields  $\beta =
I_1/I_2$, where
\begin{eqnarray*}
I_1 = \int d\eta\biggl\{
-fw_r^{(0)}\p_\eta{w_{r}^{(0)}}\left[3W_0(\eta)w_r^{(0)}+2w_i^{(1)}\right]
\nonumber \\
+ \left(\p_\eta w_{r}^{(0)}\right)^2 \left[6w_r^{(0)}w_r^{(2)} +
(w_i^{(1)})^2 - 2W_0(\eta)w_r^{(0)}w_i^{(1)}\right]
  \biggr\},
\end{eqnarray*}
and $\displaystyle I_2 = -\int
\p_\eta{W_0(\eta)}\left(w_{r}^{(0)}\right)^4 d\eta$. Finally,
using eqs.~(\ref{eq:psi22}) and (\ref{rhophi}), we can rewrite
eq.~(\ref{eq:Ell}) in the  form: $\p^2_\lm E(\sigma; {0})=2\langle
\vect{\psi}_{21}, \vect{v}\rangle = \s^2 E_{2} + o(\s^2), $ where
the  coefficient $E_{2}$   is given as $\displaystyle E_{2} =
2\int  \left(\p_\eta w_{r}^{(0)} \chi^{(1)} + \p_b
w_{r}^{(0)}\rho^{(1)}\right) d\eta $.
 Functions $\chi^{(1)}$ and $\rho^{(1)}$ can be computed numerically from the linear  equations (\ref{L-}) and (\ref{eq:Lf}), what gives an algorithm for obtaining  the coefficient $E_2$. 
For given $b$ positive value of $E_2$ corresponds to the situation
when the modes are unstable for small $\s$. \textit{Vice versa},
negative $E_2$ implies stability of the modes for small $\s$.

\section{Discussion of the results and Conclusion}

\begin{figure}
\includegraphics[width= 0.8\columnwidth]{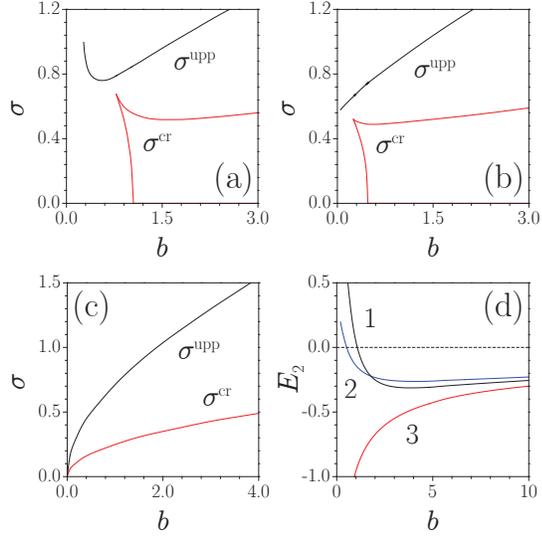} \caption{Domains of
existence and stability  of fundamental localized modes on the
plane $(b,\s)$  $\sin$-shaped~(a), $\tanh$-shaped (b), and linear
(c)  potentials $W(\eta)$ and $V(\eta)=0$. (d): The dependence of
the coefficient $E_{2}$ on $b$ in $\sin$-shaped (curve 1),
$\tanh$-shaped (curve 2), and linear (curve 3) potentials.}
\label{fig-3}
\end{figure}

Let us now turn to the results of the stability analysis of the
particular examples (see fig.~\ref{fig-3}). We start by recalling
the results for the case $V(\eta)= 0$ and $W(\eta) =  \sigma
\sin(2\eta)$ reported earlier in~\cite{AbdKartKonZez}. Physically, this case corresponds to the pure dissipative nonlinear lattice where domains with the nonlinear gain alternate with the nonlinear dissipation. This case was already discussed in~\cite{AbdKartKonZez}, and in particular it was shown that for sufficiently small $\sigma$ the nonlinear modes become stable if the propagation constant $b$   exceeds a threshold value $b^{cr}(\sigma)$. The  developed here approach based on the analysis of the Evans function 
 allows us to compute numerically $b^{cr}(0)$, as this is shown in fig.~\ref{fig-3}~(d) (the black curve 1) from which one observes that the coefficient $E_{2}$ changes its sign at $b^{cr}(0)\approx 1.05$
{which}
 corroborates results reported in the panel~(a). This critical value corresponds to narrow modes, whose widths are of order of the period of the dissipative lattice.

Next we consider potential~(\ref{tanh}) [see also
fig.~\ref{fig-1}] which is characterized by only one domain with
nonlinear gain and one domain with nonlinear dissipation. From
fig.~\ref{fig-3}~(b)  we observe that the domains of existence and
stability are similar to those obtained for the sin-shaped
potential: in particular,  there {exists} the threshold for the
stability of the modes. Quantitatively, now the stable modes
 can be broader than in the case of periodic potential:
  threshold value of propagation constant is remarkably lower than in the previous case. More specifically, using the  approach based on the Evans function   we obtain  [the blue curve in panel (d)]: $b^{cr}(0) \approx 0.49$.
 This result agrees with  fig.~\ref{fig-3}~(b) and with fig.~\ref{fig-1}~(d) where  dependence of perturbation growth rate $\RE\lm$ on $\s$ is shown.

From the results for  sin-- and tanh-- shaped  potentials we can
conjecture that (i) the most stable modes   are localized on the
scale where the dissipative potential can be approximated by the
linear function, i.e. where $W(\eta)\sim  2\sigma\eta$ on the
width of the mode; and (ii) properly introduced gain and
dissipation from the both sides of the mode enhances its
stability. These arguments readily lead to the model
\begin{eqnarray}
\label{2eta}
i \p_\xi q =-\frac 12 \p^2_\eta q  -|q|^2q  - 2i \sigma \eta |q|^2q
\end{eqnarray}
where all modes in the limit $\sigma\to 0$ should be stable. This
is indeed what happens, as one can see from fig.~\ref{fig-3}~(c):
the instability threshold  $b^{cr}(0)$ disappears completely. This
is also confirmed by analysis based on the Evans function [the red
curve 3 in the panel (d)].

Now from panel (c) we observe that both lines, i.e. upper border of existence domain
  $\sigma^{upp}$ (black curve) and the upper border of stability domain  $\sigma^{cr}$ (red curve), follow the parabolic scaling law $b\sim\sigma^2$. This law can be understood from the simple scaling arguments as follows. If $q(\xi,\eta)$ is a solution of eq.~(\ref{2eta}), then $\tilde{q}(\tilde{\xi},\tilde{\eta})=\frac 1\sigma q(\xi, \eta)$   with $\tilde{\xi} =\sigma^2\xi $ and $\tilde{\eta} =\sigma\eta $ is a solution of the complex NLS equation without any parameter:
$i \p_{\tilde{\xi}}\tilde{q} =-\frac 12 \p^2_{\tilde{\eta}}
\tilde{q}  - |\tilde{q}|^2\tilde{q}  - 2i \tilde{\eta}
|\tilde{q}|^2\tilde{q}$. This observation as well as the  fact
that a narrow mode ``feels'' only the local dissipative term,
allows one to make further conclusions about the behavior of the
curves in fig.~\ref{fig-3}. Since the mode widths tend to zero at
$b\to\infty$ (and $\sigma$ bounded), and all the examples of
$W(\eta)$  were chosen to have the same slope at $\eta=0$:
$W(\eta) \sim 2\s\eta$, the existence curves $\sigma^{upp}$ in the
panels (a) and (b)  tend at $b\to\infty$ to the parabola
$\sigma^{upp}$ shown in the panel (c). This conjecture was
supported by  numerical simulations up to $b=20$ where for the
linear, $\tanh$--, and $\sin$--shape potentials we found
$\sigma^{upp}=3.18,\ 3.20$, and $3.19$, respectively.

\begin{figure}%
\includegraphics[width=\columnwidth]{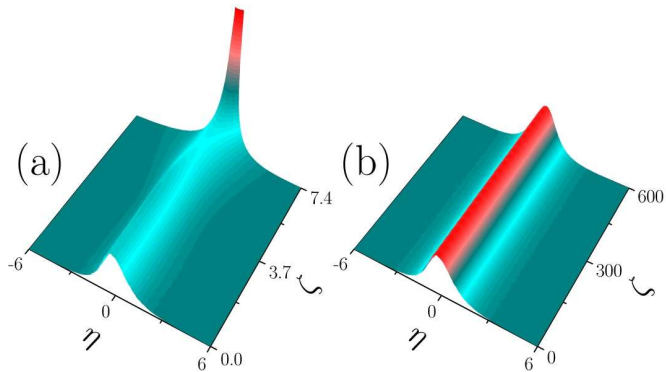}%
\caption{Propagation of initially perturbed nonlinear modes of
eq.~(\ref{2eta}) for $\s=0.5>\s^{cr}$ (a) and $\s=0.2<\s^{cr}$
(b). For both the cases $b=1$.} \label{fig-4}
\end{figure}

{In fig.~\ref{fig-4} we present two examples of evolution
of  unstable (panel a) and stable (panel b) modes. These results,
obtained by direct integration of eq.~(\ref{2eta}) [i.e. of the
particular case of the model (\ref{CNLS}) which is mostly
``exposed'' to eventual nonlinear instabilities]
confirm the results of linear stability analysis. Thus, the modes
belonging to the stability domains propagate undistorted over
indefinitely long distances, even if they are strongly perturbed
initially.}

To conclude, we have investigated fundamental modes in imaginary
\PT-symmetric nonlinear potentials. Such potentials 
allow for existence of localized modes which are stable at
least in the limit when the mode is narrow enough. The stability
was established both as the linear stability, on the basis of the
Evans-function analysis, and using direct numerical study of the
mode evolution {(notice that while the direct propagation ensures
also the nonlinear stability of the modes in  a finite domain, the
nonlinear nature of the perturbation 
may
introduce new features of the nonlinear stability of the solutions
on the whole real axis).} Although our analysis was performed for
nonlinear potentials, the established symmetry properties have
more general character and the approach can be applied also for
linear \PT-potentials, as well as to the cases where both linear
and nonlinear \PT-symmetric potentials are present. {In
the latter case   stability   of the modes may change
dramatically. 
For example, in presence of a periodic linear \PT-potential broad
small-amplitude modes are expected to be stable as long as
imaginary part of the linear  potential is below a critical value.
This situation is in contrast to the one shown in
fig.~\ref{fig-3}~(a)
where broad modes 
are unstable. As
another interesting question, we would like to mention the
exploration of asymmetric nonlinear modes similar to ones reported
in~\cite{MirMalKiv}, although their existence may require more
sophisticated nonlinearity landscapes.}

\acknowledgments DAZ and VVK were supported by FCT (Portugal)  under the
grants  No. SFRH/BPD/64835/2009 and PEst-OE/FIS/UI0618/2011.  VVK was partially supported by
the program Ac\c{c}\~{o}{e}s Integradas Luso-Espanholas No. E-27/10.

\end{document}